\documentclass[12pt]{article} 

\renewcommand{\thefootnote}{\fnsymbol{footnote}}
\usepackage{amsbsy,amssymb,latexsym,amsfonts,amsmath}
\usepackage{mathrsfs}
\usepackage{graphicx}
\usepackage{bm}
\numberwithin{equation}{section}
\newcommand{\bel}[1]{\begin{equation}\label{#1}}                     
\newcommand{\bal}[1]{\begin{eqnarray}\label{#1}}                     
\newcommand{\be}{\begin{equation}}
\newcommand{\ee}{\end{equation}}
\newcommand{\im}{\mathrm{i}}
\newcommand{\ex}{\mathrm{e}}
\newcommand{\de}{\mathrm{d}}

\newcommand{\ul}[1]{\underline{#1}}
\begin{document}
%%%%%%%%%%%%%%%%%%%%%%%%%%%%%%%%%%%%%%%%%%%%%%%%%%%%%
%%%%%%%%%%%%%%%%%%%%%%%%%%%%%%%%%%%%%%%%%%%%%%%%%%%%%
%
% title page
%
%%%%%%%%%%%%%%%%%%%%%%%%%%%%%%%%%%%%%%%%%%%%%%%%%%%%%
\begin{titlepage}
%%%%%%%%%%%%%%%%%%%%%%%%%%%%%%%%%%%%%%%%%%%%%%%%%%%%%
\begin{flushright}
\normalsize
%\filename
~~~~
NITEP 228\\
November 2024 \\
\end{flushright}
%%%%%%%%%%%%%%%%%%%%%%%%%%%%%%%%%%%%%%%%%%%%%%%%%%%%%

\vspace{15pt}

%%%%%%%%%%%%%%%%%%%%%%%%%%%%%%%%%%%%%%%%%%%%%%%%%%%%%
\begin{center}
{\LARGE Phases and triple(multiple) point: } \\
{\LARGE  critical phenomena around the AD singularity  \\
\vspace{10pt}
\LARGE }
\end{center}
%%%%%%%%%%%%%%%%%%%%%%%%%%%%%%%%%%%%%%%%%%%%%%%%%%%%%

\vspace{23pt}

%%%%%%%%%%%%%%%%%%%%%%%%%%%%%%%%%%%%%%%%%%%%%%%%%%%%%
\begin{center}
{H. Itoyama$^{a,b}$\footnote{e-mail: itoyama@omu.ac.jp},
  and  R. Yoshioka$^{a,b}$\footnote{e-mail: ryoshioka@omu.ac.jp}  }\\
%%%%%%%%%%%%%%%%%%%%%%%%%%%%%%%%%%%%%%%%%%%%%%%%%%%%%

\vspace{18pt}

%%%%%%%%%%%%%%%%%%%%%%%%%%%%%%%%%%%%%%%%%%%%%%%%%%%%%

$^a$ \textit{Nambu Yoichiro Institute of Theoretical and Experimental Physics (NITEP), Osaka Metropolitan University} \\ 

$^b$ \it Osaka Central Advanced Mathematical Institute (OCAMI), Osaka Metropolitan University\\

3-3-138, Sugimoto, Sumiyoshi-ku, Osaka, 558-8585, Japan \\

\end{center}
%%%%%%%%%%%%%%%%%%%%%%%%%%%%%%%%%%%%%%%%%%%%%%%%%%%%%

\vspace{20pt}

\begin{center}
Abstract\\
\end{center}
%%%%%%%%%%%%%%%%%%%%%%%%%%%%%%%%%%%%%%%%%%%%%%%%%%%%%

Continuing with our previous series of work, 
we present a case study of the critical phenomena around  Argyres-Douglas singularity
of ${\cal N} =2$ susy made at $(A_1, A_{4k-1} ), k =1, 2$ realized by one-unitary matrix model.
We determine the phase diagram, which is recast into LEEA of $\mathcal{N}=2$, 4d gauge theory by the  0d-4d connection. 
There are three distinct phases, each corresponding to an eigenvalue distribution with 0, 1, and 2 gaps. 
These form an entire phase diagram with a triple point.
Examining the behavior of the planar free energy, 
we show, among other things, 
 that  the transition line between 1- and 2-gap phases ending at  the triple point  
 is  the $k=2$ multicritical one.

%%%%%%%%%%%%%%%%%%%%%%%%%%%%%%%%%%%%%%%%%%%%%%%%%%%%%

\vfill

\end{titlepage}

%%%%%%%%%%%%%%%%%%%%
\renewcommand{\thefootnote}{\arabic{footnote}}
\setcounter{footnote}{0}
%%%%%%%%%%%%%%%%%%%%

%%%%%%%%%%%%%%%%%%%%%%%%%%%%%%%%%%%%%%%%%%%%%%%%%%%%
%%%%%%%%%%%%%%%%%%%%%%%%%%%%%%%%%%%%%%%%%%%%%%%%%%%%
\section{Introduction}
%%%%%%%%%%%%%%%%%%%%%%%%%%%%%%%%%%%%%%%%%%%%%%%%%%%%
%%%%%%%%%%%%%%%%%%%%%%%%%%%%%%%%%%%%%%%%%%%%%%%%%%%%

Quenching of the position dependence in  holomorphic correlation functions due to singularities 
 is a phenomenon characteristic of supersymmetric quantum field theory \cite{CDSW2002}.  
 In theory with the adjoint matter, the Schwinger-Dyson equation taking the form of Virasoro like constraints 
 clarifies the relevance of matrix models description that has come from geometric engineering 
 of string theory \cite{DV0206,DV0207,CM0209,IM0211245}\cite{Davi1990,MM1990,IMat1991N}. 
In the case of instanton expansion of $\mathcal{N}=2$ theory \cite{SW9407}\cite{Naka1999,NY0306,Nekr0206}, 
 this 0d-4d connection \cite{DV0909,IMO0911} can be demonstrated more explicitly through the AGT relation \cite{AGT0906}
as well as  the identification of regular/irregular conformal block with
matrix models whose potential contain logarithmic terms \cite{DF1984,MMS1001}
 and systematic expansion has been given \cite{IO1003}.\footnote{For a review, see, for instance, \cite{IYosh1507,Itoy2016,LeFl2006}.}

Beginning with \cite{IOYano1805,IOYano1812,IOYano2019}, 
we first observed the relevance of one-unitary matrix model with cosine potential, 
namely, the GWW model \cite{GW1980,Wadi1980} with the addition of the log term for $su(2)$ $N_f=2$ irregular conformal block, 
which is obtained from that of $N_f = 4$ by the limiting procedure \cite{IOYone1008}. 
The identification of the Argyres-Douglas singularity \cite{AD9505,APSW9511} with the critical point of this matrix model yielded the self-contained derivation of PII with accessory parameter
in the double scaling limit \cite{BK1990,GM1990NPB,GM1990PRL,DS1990}.
The conversion of the original hermitean matrix model into the unitary one has been shown to be effective in $su(n)$ (multi matrix model) extension as well \cite{IOYosh2210,IOYosh2212}
 and the AD singularity has appeared as a critical hypersurface in the space of flavour mass parameters.

In another vein of extension, the potential can be augmented up to $L$-th power of cosine
and has been seen to serve as theory space which accommodates critical singularities up to
$L$-th order and hence that for $(A_1, A_{4k-1}), k=1, \cdots, L$ \cite{IOYano2103}. 
The matching of the operator contents and their scaling dimensions on both sides has been successfully given there. 
Throughout these developments\footnote{For higher rank extension of PII and the nonperturbative effects, see \cite{IOYano1909,CIY2402}}, 
we have emphasized that we work with the partition function whose log gives free energy rather than the instanton partition function which prefers explicit specification of filling fractions as Coulomb moduli (constant background field), rendering the whole picture consistent with the developments in $N_f=4$. (See, for instance, \cite{GIL1207,MM1707}.)

In this paper, we focus on the determination of the phase diagram
in the simplest and yet non-trivial possible case: we work on $L=2$
theory space, keeping perturbations by the operator corresponding to the original matrix model action and by the cosine potential of the second order alone, leaving aside the sine potentials. 
We drop the log potential\footnote{Both sine and log are ``odd" operators.} as well: 
perturbation by this operator has been identified to correspond to the mass operator 
 (actually the difference in the two flavour masses \cite{IOYano1805,IOYano1812,IOYano2019})
 and will not be discussed in this paper. 
 With this setup containing the two couplings denoted by $(\lambda,\tau)$ only, 
 the matrix model coincides with the one first considered by Mandal \cite{Mand1990}.\footnote{For other solvable extensions and the phase diagrams, see \cite{ST2102}.}

In this subspace $\mathcal{S}$ of $L=2$ theory space that represents an extension of $N_f=2$ $su(2)$ asymptotically free theory \cite{NR1207}, 
we study critical phenomena associated with the AD singularity of $(A_1, A_{4k-1}), k = 1, 2$ type. 
After giving some preliminaries in the next section,  we present, in section three, 
complete determination of the phase diagram in the planar limit.
In the upper half plane of $(\lambda,\tau)$, we give four critical lines that separate two out of the three different phases denoted by 2 gap, 1 gap and 0 gap ones.\footnote{In \cite{Mand1990}, the presence of $0, 1, 2$ gap phases was noted but the phase separation lines were not given.}
These lines are the AD singularities in $\mathcal{S}$.
We find that two out of three $k=1$ lines and $k=2$ line merge smoothly at a point, forming a triple point. 
The $k=2$ singularity is a full-fledged line rather than a point contrary to  naive expectation one may have.

\section{Some preliminaries}

%%%%%%%%%%%%%%%%%%%%%%%%%%%%%%%%%%%%%%%%%%%%%%%%%%%%
\subsection{notation}\label{notaion}
%%%%%%%%%%%%%%%%%%%%%%%%%%%%%%%%%%%%%%%%%%%%%%%%%%%%
While we work eventually in the notation set up in the next section to determine the phase diagram, 
let us begin with the notation of \cite{CIY2402} and that of the earlier papers.
The unitary matrix model with logarithmic interaction can be represented by 
\begin{equation}\label{Z_U}
Z_{U(N)} = \frac{1}{\mathrm{Vol}(N)} \int [\de U] \exp\Big(
 \mathrm{Tr}\, W_U(U) 
\Big), 
\end{equation}
\begin{equation}\label{W_U}
W_U(w) = -q_{02} \left(w + \frac{1}{w} \right) + M \log w. 
\end{equation}

In the original $\beta$ deformed matrix model for the $N_f=2$, 
there are four net 0d parameters aside from $q_{02}$ under one  constraint. 
The number reduces to two by setting $\beta=1$ and not specifying the filling fraction. 
Namely, 
\begin{equation}
\alpha_{1+2},~~ \alpha_{3+4},~~ N , 
\end{equation}
under the constraint: 
\begin{equation}
\alpha_{1+2} + \alpha_{3+4}+ 2N = 0. 
\end{equation}

Likewise, there are four 4d parameters aside from $\Lambda_2$, the 4d scale parameter. 
This number reduces to two by setting $\frac{\epsilon_1}{g_s} = 1$ and not referring to $a$: 
\begin{equation}
 \frac{m_1}{g_s},~ \frac{m_2}{g_s}.
\end{equation}

The 0d-4d dictionary is 
\begin{equation}\label{dictionary}
\begin{split}
&\alpha_{1+2} = \frac{2m_2}{g_s},\\
&\alpha_{3+4} = \frac{2m_1}{g_s}, \\
& N = - \frac{m_1+m_2}{g_s}.
\end{split}
\end{equation}

Separately, the parameter for the 4d instanton expansion is 
\begin{equation}
q_{02} = \frac{\Lambda_2}{2g_s} \equiv \frac{1}{2}\frac{1}{\underline{g}_s}. 
\end{equation}
and $M$ in eq.  \eqref{W_U} is related to the 4d parameter by 
\begin{equation}\label{M}
M \equiv \alpha_{3+4} + N = \frac{m_1-m_2}{g_s}. 
\end{equation}
The 4d meaning of $N$, $M$ in eqs. \eqref{Z_U}, \eqref{W_U} has become manifest. 

For large $N$ analysis, it is more preferable to recast eq. \eqref{W_U} into 
\begin{equation}
W_N(w) = \frac{N}{\tilde{S}} \left\{ -\frac{1}{2} \left( w+ \frac{1}{w} \right) + \zeta \log w\right\}. 
\end{equation}
Here, we have chosen to regard the two parameters as
\begin{equation}
\tilde{S} \equiv N \underline{g}_s = -\frac{m_1+m_2}{\Lambda_2},  
\end{equation}
and
\begin{equation} 
\zeta \equiv M \underline{g}_s = \frac{m_1-m_2}{\Lambda_2} , 
\end{equation}
while $1/N$ is the expansion parameter. 

As stated in introduction, we consider the following extension of the above by the even interactions, 
focusing on $p_{\rm max} = 2$, 
\begin{equation}\label{ZU}
 Z_U = \frac{1}{N!} \left( \prod_{j=1}^N \oint \frac{dz_j}{2 \pi z_j}\right) \Delta(z) \Delta(z^{-1}) 
 \exp \left( \sum_j W(z_j) \right), 
\end{equation}
\begin{equation}
 W(z) = -\frac{1}{2\ul{g}_s} \sum_{p=1}^{p_{\rm max}} \frac{g_p}{p}\left( z^p + \frac{1}{z^p} \right). 
\end{equation}

%%%%%%%%%%%%%%%%%%%%%%%%%%%%%%%%%%%%%%%%%%%%%%%%%%%%
%%%%%%%%%%%%%%%%%%%%%%%%%%%%%%%%%%%%%%%%%%%%%%%%%%%%
\section{Determination of phases from the planar free energy}
%%%%%%%%%%%%%%%%%%%%%%%%%%%%%%%%%%%%%%%%%%%%%%%%%%%%
%%%%%%%%%%%%%%%%%%%%%%%%%%%%%%%%%%%%%%%%%%%%%%%%%%%%
Following the outline in the introduction, we will determine the phase diagram  at the planar level completely. 
 
%%%%%%%%%%%%%%%%%%%%%%%%%%%%%%%%%%%%%%%%%%%%%%%%%%%%
\subsection{notation}
%%%%%%%%%%%%%%%%%%%%%%%%%%%%%%%%%%%%%%%%%%%%%%%%%%%%
As we make a direct contact with the original work of \cite{Mand1990}, we first convert our notation. 
Let us put the integral variables as $z_j = e^{\im \alpha_j}, -\pi \leq \alpha \leq \pi$ and  
 introduce two coupling constants, 
\begin{equation}
 \lambda = - \frac{2N \ul{g}_s}{g_1}, ~~~~~
 \tau = \frac{g_2}{2g_1}.
\end{equation}
Ignoring the overall coefficient $\im^N/N! (2\pi)^N$,  eq. \eqref{ZU} becomes 
\begin{equation}
 Z = \int \prod_{i=1}^N \de \alpha_i 
 \prod_{i\neq j}   |\ex^{\im \alpha_i} - \ex^{\im \alpha_j}|
  \exp \left\{\frac{N}{\lambda} \sum_{i} V(\alpha_i) \right\}, 
\end{equation}
\begin{equation}
 (-V(\alpha)) = -2 ( \cos \alpha + \tau \cos 2\alpha)  \equiv  2 U(\alpha),
\end{equation}
where
\begin{equation}
 U(\alpha) = -\cos \alpha - \tau \cos 2\alpha. %~~~~~ \tau \equiv \frac{a_2}{a_1}. 
\end{equation}
Since $U(\alpha)$ is an even function in $\alpha$, we pay our attention to $0 \leq \alpha \leq \pi$.  

%%%%%%%%%%%%%%%%%%%%%%%%%%%%%%%%%%%%%%%%%%%%%%%%%%%%
\subsection{extrema of the classical potential}
%%%%%%%%%%%%%%%%%%%%%%%%%%%%%%%%%%%%%%%%%%%%%%%%%%%%
As is well-known, the eigenvalue distributions of matrix models in general are 
 qualitatively understood by the local minima of the classical potential combined 
 with the 2d Coulomb repulsion among the eigenvalues given by log of the Vandermonde  determinant. 
In the large $N$ limit, the distributions become continuous to form cuts 
 (or gaps in the cases of unitary matrix models).  
 We will, therefore, briefly describe the change of the potential with $\tau$.  

\begin{figure}[h]
\centering
\begin{minipage}[b]{0.3\columnwidth}
    \centering
    \includegraphics[width=0.9\columnwidth]{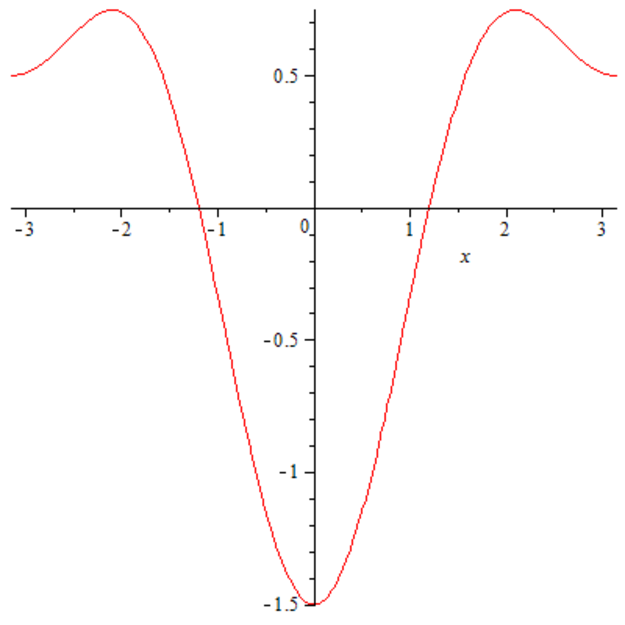}
    \caption{$\tau$= $\frac{1}{2}$,  
    $\alpha_0   = \frac{2\pi}{3}$.}
    \label{fig:U(1/2)}
\end{minipage}
\begin{minipage}[b]{0.3\columnwidth}
    \centering
    \includegraphics[width=0.9\columnwidth]{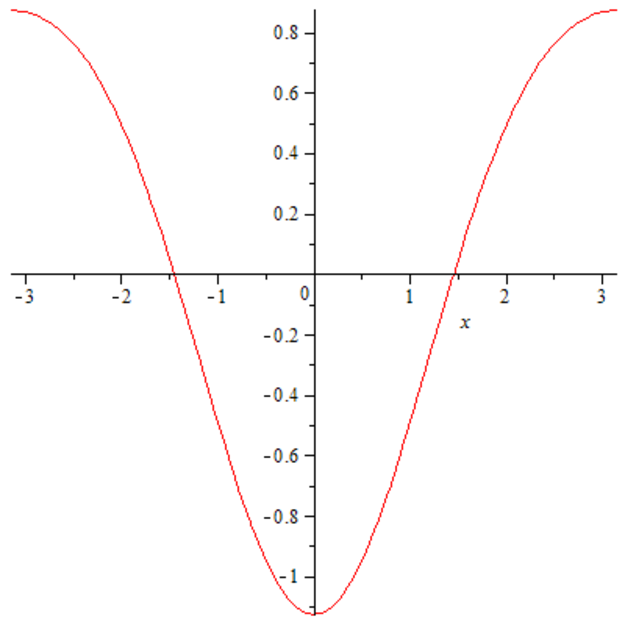}
    \caption{$\tau = \frac{1}{8}$.}
    \label{fig:U(1/8)}
\end{minipage}
\begin{minipage}[b]{0.3\columnwidth}
    \centering
    \includegraphics[width=0.9\columnwidth]{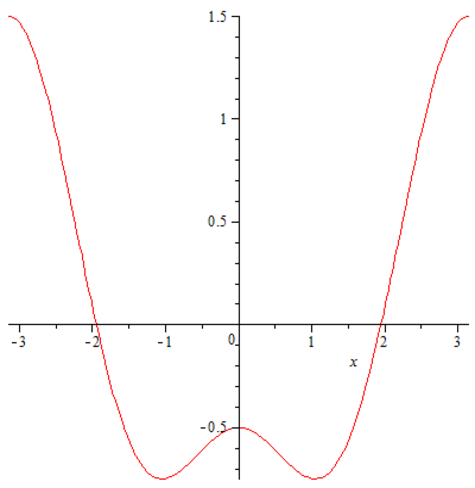}
    \caption{$\tau = -\frac{1}{2}$,  $\alpha_0= \frac{\pi}{3}$.}
    \label{fig:U(-1/2)}
\end{minipage}
\end{figure}

The extrema of $U(\alpha)$ are given by 
\begin{equation}
 U'(\alpha) = \sin \alpha (1+4\tau \cos \alpha) = 0. 
\end{equation}
We will see that $\alpha=0$, $\pi$ are always extrema 
 while $\alpha_0 \equiv \cos^{-1}(-\frac{1}{4\tau})$ is an extremum only when $|\tau| \leq \frac{1}{4}$.
We  conclude that 

i) $\tau > \frac{1}{4}$;~
 two local minima at $\alpha = 0$ and $\pi$ and two local maxima at $\alpha = \pm \alpha_0$.    
 
ii) $|\tau| \leq  \frac{1}{4}$;~ 
 %two local minima at $\alpha = 0$, $\pi$ only. 
 a local minimum at $\alpha=0$ and a local maximum at $\alpha = \pi$ only.  
 
iii) $\tau < -\frac{1}{4}$; ~
 two local maxima at $\alpha = 0$, $\pi$ and two local minima at $\alpha = \pm \alpha_0$. \\
Fig. \ref{fig:U(1/2)}, Fig. \ref{fig:U(1/8)} and Fig. \ref{fig:U(-1/2)} illustrate these cases i), ii) and iii) respectively.
%%%%%%%%%%%%%%%%%%%%%%%%%%%%%%%%%%
%%%%%%%%%%%%%%%%%%%%%%%%%%%%%%%%%%

The behavior of the eigenvalue distribution can be examined in part from the shape of potential  classically. 
However, only its $\tau$-dependence is observed. 
In order to determine the actual eigenvalue distribution, we have to  pay attention to the $\lambda$-dependence as well. 
Fortunately, it is easy to incorporate the $\lambda$-dependence as $\lambda$ behaves as 
 a parameter that measures repulsion between eigenvalues.
For large $\lambda$, the repulsion also becomes large and the gaps should shrink in any of these cases.  
As $\lambda$ decreases, the shape of the potential begins to affect the eigenvalue distribution.
Let us first understand the qualitative behavior of the distribution  in each case. 

\noindent
i)
If $\lambda$ decreases, the eigenvalues accumulate around the local minima $\alpha = 0$ and $\pi$.  This situation corresponds to two-gap phase. 
At a sufficiently small $\lambda$, all eigenvalues gather around the minimum $\alpha = 0$.
Hence, one-gap phase for small $\lambda$ and $0 \to 2 \to 1$ transition is expected. 

\noindent
 ii) 
In this case, the minimum is only at $\alpha=0$ and $0 \to 1$ transition is expected.

\noindent
 iii) 
There exists a local maximum at $\alpha=0$. 
 If $\lambda$ is large enough, the eigenvalues will overcome that barrier.
This situation corresponds to one-gap phase. 
As $\lambda$ gets smaller, the eigenvalues are distributed around each of the two minima. 
Hence, two-gap phase for small $\lambda$ and $0 \to 1 \to 2$ transition is expected.

%%%%%%%%%%%%%%%%%%%%%%%%%%%%%%%%%%%%%%%%%%%%%%%%%%%%
\subsection{separation lines and the triple point}
%%%%%%%%%%%%%%%%%%%%%%%%%%%%%%%%%%%%%%%%%%%%%%%%%%%%
In this section, we derive the transition lines which separate each phase and 
 determines the end points of gaps on these lines. 
We will see that there  exists a triple point in the phase diagram. 

\subsubsection{$0 \to 1$ and $0 \to 2$}
The spectral density for the 0-gap solution is given by \cite{GW1980,Mand1990}
\begin{equation}\label{density}
 \rho(\alpha)  = \frac{1}{2\pi} 
 \left( 1 + \frac{2}{\lambda} \nu(y) \right),~~~
 y = \sin^2 \frac{\alpha}{2}, 
\end{equation}
\begin{align} \label{nu}
 \nu(y)  &= 1 + 2\tau - y(2+16\tau) + 16\tau y^2. 
\end{align}
In the large $\lambda$ limit, $\rho(y) \sim \frac{1}{2\pi}$, 
 namely,  the eigenvalues fill the entire unit circle.  
 
Transitions should take place when a point $\alpha = \alpha_0$ that satisfies $\rho^{(0)}(\alpha_0) =0$ appears. 
 As $\lambda$ decreases, eq.\eqref{nu} can take negative values and, at $\alpha=\alpha_0$, 
\begin{equation} \label{rho0=0}
 1+\frac{2}{\lambda}\nu(y_0) = 0,~~~~~ y_0 = \sin^2\frac{\alpha_0}{2}. 
\end{equation}
If $\lambda$ further decreases, the positivity of the spectral density is violated.  
Hence there exists a lower bound of $\lambda$, which is given by eq.\eqref{rho0=0}, 
\begin{equation} 
 \ul{\lambda}^{(0)}  =  -2 (1+ 2 \tau - y_0 (2+16 \tau) + 16 \tau y_0^2 ). 
\end{equation}

\begin{figure}[h]
\centering
    \includegraphics[width=0.5\columnwidth]{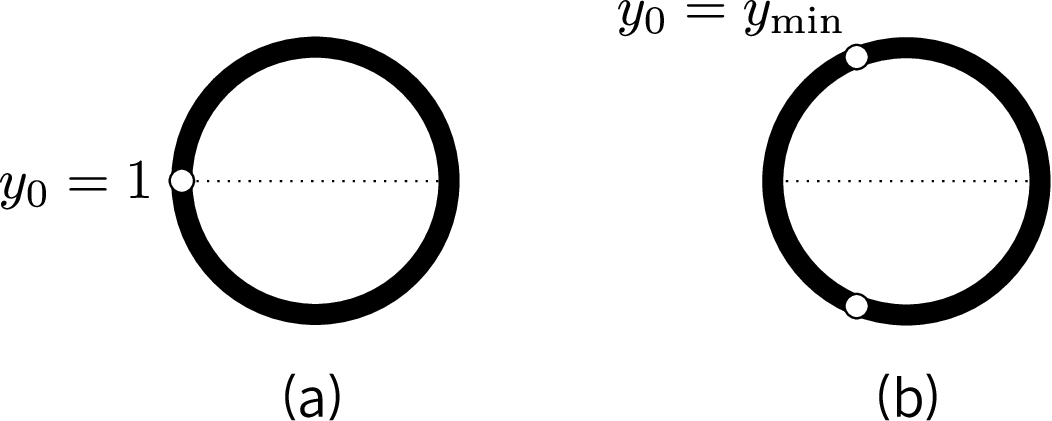}
    \caption{(a) $\tau \leq \frac{1}{8}$, (b) $\tau > \frac{1}{8}$}
    \label{Fig:012transition}
\end{figure}

Let us determine $y_0$ which is a point such that $\nu(y)$ takes the negative minimum value 
 for  $0 \leq y  \leq 1$.  
Eq. \eqref{nu} is a quadratic equation in $y$ and takes the minimum value at  
\begin{equation} 
 y_{\rm min} = \frac{1+8\tau}{16 \tau}.
\end{equation}
If $\tau \leq \frac{1}{8}$, then $y_{\rm min} \geq 1$ and, 
 therefore,  $y_0 =1$  (i.e. $\alpha_0 = \pi$). 
The point $y_0=1$ is shown in Fig.\ref{Fig:012transition}(a). 
In this case, we obtain 
\begin{equation}\label{transition:01}
 \lambda_c^{(01)}(\tau) = \ul{\lambda}^{(0)}(y_0=1)  = -4\tau + 2,   
\end{equation}
which corresponds to  the transition line separating zero-gap and one-gap in the phase diagram.   

If $\tau > \frac{1}{8}$, $y_0 = y_{\rm min}(<1)$. 
From the symmetry of the eigenvalue distribution shown in Fig. \ref{Fig:012transition}(b),
 we see that two gaps open in this case and we obtain 
\begin{equation}\label{transition:02}
 \lambda_c^{(02)}(\tau) = \ul{\lambda}^{(0)}(y_0=y_{\rm min}) 
 = 2a_1\left( 2 \tau + \frac{1}{16 \tau} \right), 
\end{equation}
which is transition line between zero-gap and two-gap. 
%%%%%%%%%%%%%%%%%%%%%%%%%%%%%%%%%%
%%%%%%%%%%%%%%%%%%%%%%%%%%%%%%%%%%

\subsubsection{$1 \to 0$ and $1 \to 2$}
\begin{figure}[h]
\centering
    \includegraphics[width=0.15\columnwidth]{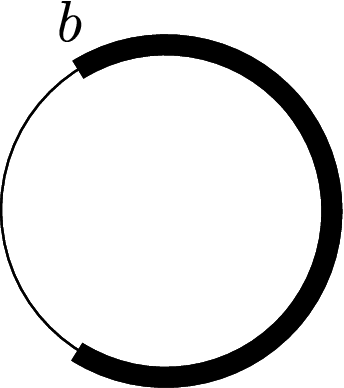}
    \caption{1-gap solution. The eigenvalue distribution is depicted by the bold line.}
    \label{Fig:1-gap}
\end{figure}

The 1-gap solution is given by 
\begin{equation}
 \rho(\alpha) = \frac{1}{2\pi} g(\alpha) \cos \frac{\alpha}{2} \sqrt{b - \sin^2 \frac{\alpha}{2}}, 
\end{equation}
where 
\begin{align}
&g(\alpha) = g_0 + g_1 \cos \alpha, \\
&g_0 = \frac{4}{\lambda} (1-4 \tau b), \\ 
&g_1 = \frac{16 \tau}{\lambda}, \\ 
& b \equiv \sin^2 \frac{\alpha_c}{2} = \frac{(1+4\tau) -\sqrt{(1+4\tau)^2-12 \tau \lambda} }{12\tau}.
\end{align}
The parameter $\alpha_c $ is the end point of gap and is also equivalently that of  eigenvalues.
The eigenvalue distribution is shown in Fig. \ref{Fig:1-gap}.
%%%%%%%%%%%%%%%%%%%%%%%%%%%%%%%%%%
%%%%%%%%%%%%%%%%%%%%%%%%%%%%%%%%%%

As $\lambda$ increases, the eigenvalue distribution grows while the gap shrinks.
The gap is finally shrunk to a point and the $1 \to 0$ transition occurs.  
This point should be $b=1$ (i.e. $\alpha_c = \pi$)  which gives the relation \eqref{transition:01} again
 for $\tau \leq \frac{1}{8}$. 
In the case of $\tau > \frac{1}{8}$,  however,  on the line given by \eqref{transition:01}, 
\begin{equation}
 b = \frac{1-2\tau}{6\tau} < 1, 
\end{equation} 
which  means that the gap is not shrunk to a point on the line. 

Actually, there is another restriction on $\lambda$. 
Since $b$ has to be a real number, we see that $\lambda$ has an upper bound, 
\begin{equation}
 \lambda \leq \frac{(1+4 \tau)^2}{12 \tau} \equiv \overline{\lambda}^{(1)}(\tau) .
\end{equation}
In the case of $\tau < \frac{1}{8}$,  $\lambda^{(01)}_c < \overline{\lambda}^{(1)}$ is satisfied and, therefore, the $1 \to 0$ transition occurs before $\lambda$ increases up to the upper bound $\overline{\lambda}^{(1)}$. 
In the case of $\tau \geq \frac{1}{8}$,  
$\overline{\lambda}^{(1)}$ is the genuine upper bound. 
At this point the one-gap solution fails, calling for another kind of transition which cannot be identified with the 1 $\to 0$ transition. 
Instead, this transition ought to be the $1 \to 2$ transition and 
 the upper bound  itself becomes the separation line,     
\begin{equation}\label{transition:12}
\lambda^{(12)}_c(\tau) = \overline{\lambda}^{(1)}(\tau) = \frac{(1+4 \tau)^2}{12 \tau}. 
\end{equation}
The end point of the gap on the $1\to 2$ transition line is given by 
\begin{equation}
 b^{(12)}_c = \frac{1+4\tau}{12 \tau}.
\end{equation}

There exists another $1 \to 2$ transition  in the case of $\tau \equiv - \tilde{\tau} <0$. 
The end point of the gap is
 \begin{equation}\label{b:tautilde}
  b = \frac{(4 \tilde{\tau}-1) +\sqrt{(1-4 \tilde{\tau})^2 + 12  \tilde{\tau} \lambda} }{12 \tilde{\tau}}. 
 \end{equation}
 As $\lambda$ decreases, so does $b$, and, however,
 the condition $\rho(\alpha) \geq 0$ gives the lower bound of $b$,  for $\tilde{\tau}> \frac{1}{4}$,
\begin{equation}\label{btilde}
 b \geq \tilde{b}^{(12)}_c \equiv\frac{4\tilde{\tau}-1}{4\tilde{\tau}} = \frac{4\tau+1}{4\tau} > 0. 
\end{equation}
Hence it is expected that the phase transition occurs at the lower bound. 
This transition has to be $1 \to 2$ because the gap grows. 
The transition line can be determined by  $b = \tilde{b}_c^{(12)}$,     
\begin{equation}
 \tilde{\lambda}_c^{(12)}(\tau) = \frac{(4\tilde{\tau}-1)^2}{4 \tilde{\tau}} = -\frac{(4{\tau}+1)^2}{4\tau}.
\end{equation}

\subsubsection{$2 \to 0$ and $2 \to 1$}
\begin{figure}[h]
\centering
    \includegraphics[width=0.15\columnwidth]{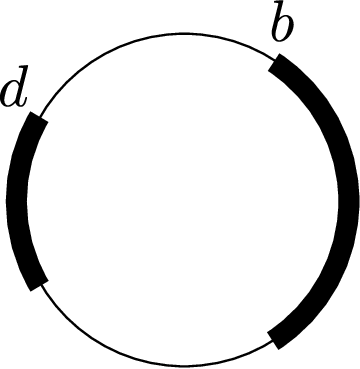}
    \caption{2-gap solution.}
    \label{Fig:2-gap}
\end{figure}

The 2-gap solution is given by 
\begin{equation}\label{solution:2-gap}
\rho^{(2)}(\alpha) = \frac{1}{2\pi} g(\alpha)  \sqrt{\left(b - \sin^2 \frac{\alpha}{2}\right)\left(d - \sin^2 \frac{\alpha}{2}\right)}, 
\end{equation}
where 
\begin{align}
&g(\alpha) = g_0 + g_1 \cos \alpha, \\
&g_0 = \frac{4}{\lambda} (1-4\tau (b+d-1)), \\
&g_1 = \frac{16\tau}{\lambda}. 
\end{align}
The end points of gaps $b$ and $d$ satisfy 
\begin{equation}
\lambda = 2(b+d-1) + \tau (-8 (b+d-1)^2 + 4(1+b-d)(1-b+d)).
\end{equation}
The eigenvalue distribution is shown in Fig. \ref{Fig:2-gap}.
%%%%%%%%%%%%%%%%%%%%%%%%%%%%%%%%%%
%%%%%%%%%%%%%%%%%%%%%%%%%%%%%%%%%%

For the $2 \to 0$ transition,  all gaps are closed and, 
 therefore, the end points of the gap have to coincide,  namely, $b=d$.
Since the value of $\lambda$ should be the same as \eqref{transition:02} at this transition, 
we obtain 
\begin{equation}
 b^{(02)}_c = d^{(02)}_c = \frac{1+8\tau}{16\tau}. 
\end{equation}
 
The $2 \to 1$ transition is expected to occur
 when $b = 0$ or $d = 1$ and the two gaps are connected to one,
 in accordance with the eigenvalue distribution shown in Fig. \ref{Fig:2-gap}. 
In order for a successful transition to the 1-gap solution shown in Fig \ref{Fig:1-gap}, 
 the connected point should be $d = 1$.
In addition, by demanding that  the eigenvalue density is positive,  
 we obtain the transition line \eqref{transition:12} again.

\subsubsection{phase diagram}
\begin{figure}[h]
\centering
    \includegraphics[scale=0.6]{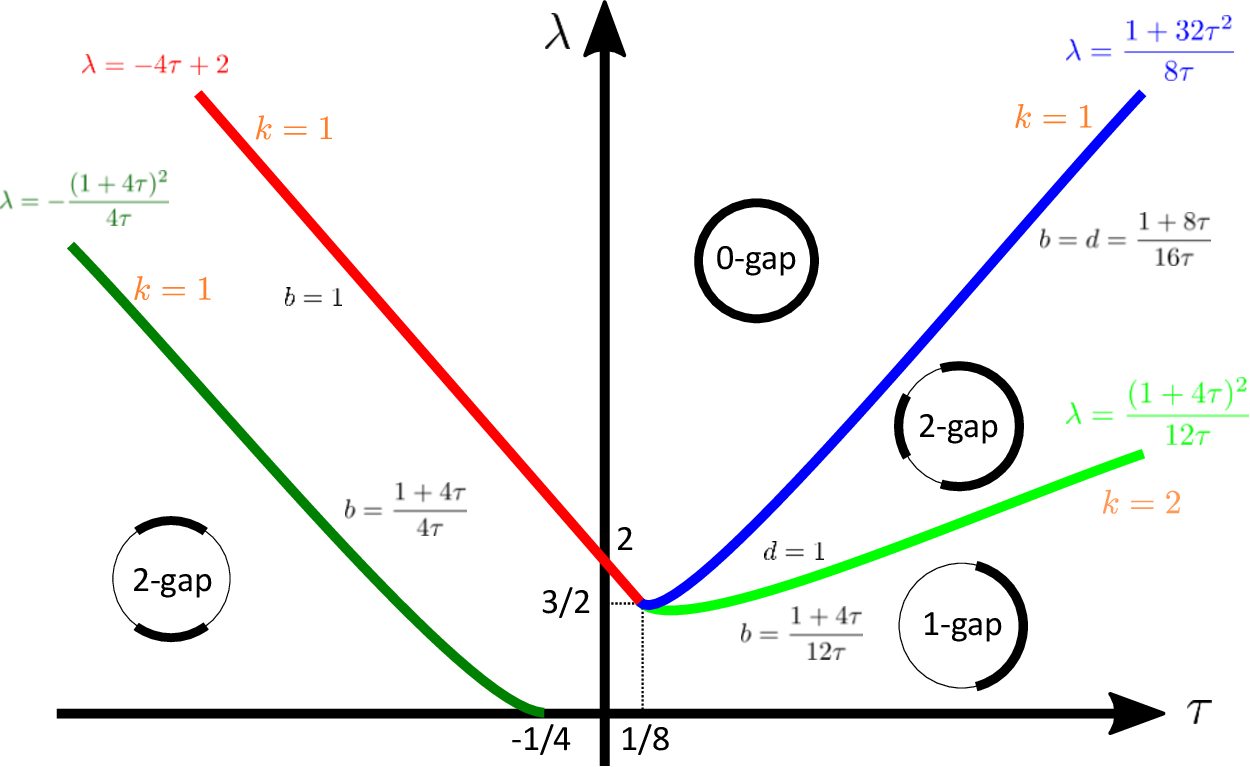}
    \caption{phase diagram}
    \label{Fig:PD}
\end{figure}

Summarizing the above results, the phase diagram can be depicted as shown in Fig. \ref{Fig:PD}. 
%%%%%%%%%%%%%%%%%%%%%%%%%%%%%%%%%%
%%%%%%%%%%%%%%%%%%%%%%%%%%%%%%%%%%

It is easy to see that the transition lines are smoothly  connected at $\tau= \frac{1}{8}$,
\begin{equation}
\begin{split}
 &\lambda_c^{(01)} (1/8)  =  \lambda_c^{(02)}(1/8)=  \lambda_c^{(12)}(1/8) = \frac{3}{2}, \cr 
 &\lambda_c^{(01)'} (1/8)  =  \lambda_c^{(02)'}(1/8)=  \lambda_c^{(12)'}(1/8) = -4,
\end{split}
\end{equation}
where $^{\prime}$ means the $\tau$ derivative.
The point  $(\tau,\lambda) = (\frac{1}{8}, \frac{3}{2})$ is the triple point. 

A similar phase diagram is obtained even when $\lambda$ is negative and 
 the entire phase diagram is symmetric with respect to the origin.

%%%%%%%%%%%%%%%%%%%%%%%%%%%%%%%%%%%%%%%%%%%%%%%%%%%%
\subsection{planar free  energy}
%%%%%%%%%%%%%%%%%%%%%%%%%%%%%%%%%%%%%%%%%%%%%%%%%%%%
The planar free energy is given by 
\begin{equation}
 F_0 = \lim_{N \to \infty} \frac{\ln Z}{N^2} = \frac{1}{\lambda} \int d\alpha \rho(\alpha) \left[
  \frac{V(\alpha)+V(0)}{2} + \lambda \ln \left|2\sin\frac{\alpha}{2}\right|
 \right].
\end{equation}
Since  $\rho(\alpha)$ is already obtained for each phase, 
 it is possible to calculate the free energy exactly.

For the zero-gap solution, it is easy to obtain  
\begin{equation}
 F^{\text{(0-gap)}}_0 = \frac{1}{\lambda^2} (1 + 2 \tau^2). 
\end{equation}

For the one-gap solution, we obtain 
\begin{align}
 F^{\text{(1-gap)}}_0 = &-\frac{1}{\lambda^2} \biggl\{
  18 \tau^2 b^4 -b^3(40 \tau^2 +10 \tau) + b^2(40 \tau^2 + 32\tau + 1) -b(16\tau^2 + 20 \tau + 4)
  \biggr\} \nonumber \\
  & -\frac{1}{\lambda} \biggl\{
  (6\tau b^2 -b(1+4\tau))\ln b  +(\ln 2)(-12\tau b^2 +b(2+ 8\tau)) - 3 \tau b^2 + b(1+4\tau)
  \biggr\} \nonumber \\
  &+ \ln 2.
\end{align}

In order to investigate the multicriticality at each phase transition, let us define the susceptibility for $\lambda$ by 
\begin{equation}
\chi = \frac{\partial^2 F_0}{\partial \lambda^2} .
\end{equation}
Around the $k$-th multicritical point $\lambda_C$, the susceptibility behaves as 
\begin{equation}
 \chi \sim (\lambda - \lambda_c)^{\frac{1}{k}}. 
\end{equation}

The zero-gap susceptibility can be expand around the transition lines as,  
\begin{equation}\label{sus:0}
\chi_{\text{0-gap}} = \frac{\partial^2 F_0^{\text{(0-gap)}}}{\partial \lambda^2}  = \frac{6(1+2\tau^2)}{\lambda_c^4} -\frac{24(1+2\tau^2)}{\lambda_c^5} { (\lambda-\lambda_c)}+\cdots,
\end{equation}
where $\lambda_c =\lambda_c^{(01)}$ or $\lambda_c = \lambda_c^{(02)}$, 
 and $\lambda_c^{(01)}$ is the $0 \to 1$ transition line  and $\lambda_c^{(02)}$ is the  $0 \to 2$ transition line, respectively.
 
Similarly, the one-gap susceptibility is, around the $1 \to 0$ transition line,  
\begin{equation}\label{sus:10}
\chi_{\text{1-gap}}^{(1\to 0)} =\frac{\partial^2 F_0^{\text{(1-gap)}}}{\partial \lambda^2}  = \frac{6(1+2\tau^2)}{(\lambda_c^{(01)})^4} - \frac{4(88\tau^3+42\tau-5)}{(1-8\tau)(\lambda_c^{(01)})^5} { (\lambda^{(01)}_c-\lambda)}+\cdots. 
\end{equation}
The constant term is coincidence with \eqref{sus:0}, but not the first order term, which implies that the phase transition is  of third order. 
On the $1 \to 2$ transition line $\lambda = \lambda_c^{(12)}(\tau)$, we obtain
\begin{equation}\label{sus:12}
\chi_{\text{1-gap}}^{(1\to 2)} = \frac{\partial^2 F_0^{\text{(1-gap)}}}{\partial \lambda^2}  
 = \frac{36 \tau^2 (176\tau^2+184\tau-1)}{(1+4\tau)^6} - \frac{288 \sqrt{3} \tau^{5/2}}{(1+4\tau)^5}    (\lambda^{(12)}_c-\lambda)^{1/2}+\cdots, 
\end{equation}
which shows that the transition line $\lambda = \lambda_c^{(12)}(\tau)$ is the $k=2$ multicritical one. 
In the vicinity of another $1 \to 2$ transition line $\lambda = \tilde{\lambda}_c^{(12)}(\tau)$ 
 at negative $\tau$, however, 
we obtain
\begin{equation}
 \tilde{\chi}_{\text{1-gap}}^{(1\to 2)} 
 = \frac{\tau^2(1392 \tau^2 -184 \tau + 197)}{2 (1+4\tau)^6} 
 + \frac{\tau^2 (68512 \tau^3 + 224 \tau^2 +8374 \tau + 81)}{4(1+4\tau)^8} (\lambda-\tilde{\lambda}_c^{(12)}) + \cdots, 
\end{equation}
and the multicritical behavior is not exhibited. 

The value of $k$ on each transition lines is also summarized in Fig. \ref{Fig:PD}.

%%%%%%%%%%%%%%%%%%%%%%%%%%%%%%%%%%%%%%%%%%%%%%%%%%%%%
%%%%%%%%%%%%%%%%%%%%%%%%%%%%%%%%%%%%%%%%%%%%%%%%%%%%%
\section*{Acknowledgments}
We thank Takahiro Nishinaka for helpful discussions.
The work of H.I. is supported in part by JSPS KAKENHI (23K03393, 23K03394).
%%%%%%%%%%%%%%%%%%%%%%%%%%%%%%%%%%%%%%%%%%%%%%%%%%%%%
%%%%%%%%%%%%%%%%%%%%%%%%%%%%%%%%%%%%%%%%%%%%%%%%%%%%%

%%%%%%%%%%%%%%%%%%%%%%%%%%%%%%%%%%%%%%%%%
%\bibliographystyle{arxiv}
%\bibliography{multicritical}
%%%%%%%%%%%%%%%%%%%%%%%%%%%%%%%%%%%%%%%%%

%%%%%%%%%%%%%%%%%%%%%%%%%%%%%%%%%%%%%%%%%%%%%%%%%%%%%

%%%%%%%%%%%%%%%%%%%%%%%%%%%%%%%%%%%%%%%%%%%%%%%%%%%%%
 
%%%%%%%%%%%%%%%%%%%%%%%%%%%%%%%%%%%%%%%%%%%%%%%%%%%%%
%%%%%%%%%%%%%%%%%%%%%%%%%%%%%%%%%%%%%%%%%%%%%%%%%%%%%
\end{document}